# Tunnel Magnetoresistance and Temperature Related Effects in Magnetic Tunnel Junctions with Embedded Nanoparticles

Arthur Useinov[*,†,‡,§] and Chih-Huang Lai[§,¶]

[*]*Department of Physics, National Tsing Hua University,
101, Sec. 2, KungFu Rd, HsinChu, 30013, Taiwan*

[†]*Department of Materials Science and Engineering,
National Tsing Hua University,
101, Sec. 2, KungFu Rd, HsinChu, 30013, Taiwan*

[‡]*Solid State Physics Department, Kazan Federal University,
18 Kremlyovskaya str., Kazan, 420008, Russia*

[§]*arthur.useinov@gmail.com,*
[¶]*chlai@mx.nthu.edu.tw*

Temperature dependence of the tunnel magnetoresistance (TMR) was calculated in range of the quantum-ballistic model in the magnetic tunnel junction (MTJ) with embedded nanoparticles (NPs). The electron tunnel transport through NP was simulated in range of double barrier approach, which was integrated into the model of the magnetic point-like contact. The resonant TMR conditions and temperature impact were explored in detail. Moreover, the possible reasons of the temperature induced resonant conditions were discussed in the range of the lead-tunneling cell (TC)-lead model near Kondo temperature. We also found that redistribution of the voltage drop becomes crucial in this model. Furthermore, the direct tunneling plays the dominant role and cannot be omitted in the quantum systems with the total tunneling thickness up to 5-6 nm. Hence, Coulomb blockade model cannot explain Kondo-induced TMR anomalies in nanometer-sized tunnel junctions.

## 1. Introduction

Theoretical modelling of the conducting properties of the nanosized quantum objects (QOs) is one of typical and important scientific problem. In this paper, we specified temperature-related effects of the TMR behavior in magnetic tunnel junction with embedded nanoparticles (npMTJ) in range of developed quantum-ballistic approach.[1,2] It is important to note that widely used consecutive and related Coulomb blockade models in literature for similar systems cannot achieve the limit of the ballistic approach due to its basic definition. For example, exploring the tunnel magnetoresistance (TMR) anomalies in magnetic tunnel junctions (MTJs) with embedded nanoparticles (NPs)[3,4] the results were explained by Coulomb blockade effects in range of the Kondo-assisted cotunneling and sequential tunneling regimes of the conductance, but direct tunneling was omitted.

In this work, we explained peak-like TMR behaviors in range of only direct tunneling and its temperature dependences based on experimental data.[3,4] Simulations were performed for electron tunneling through the insulating layer with embedded non-magnetic NPs within an approach of the single and double barrier subsystems connected in parallel. The transport problem was developed in terms of electron wave functions and quantum transmission, which determines the superposition of the quantum states for magnetic layers, barriers and NPs. The key parameter is a wave-vector value $k^{NP}$ in NP. The double barrier approach[5,6] was integrated into *a*-spot model of the magnetic point-like contact.

The developed technique is very promising and can be conjugated with the first principal calculations[7] and consistent with simulations of the planar tunnel junctions.[8,9]

## 2. *a*-Spot Model of the Magnetic Point-Contact

One can imagine metallic point-like contact, Fig. 1(a). The charge current of the metallic point-contact with orifice cross-section can be calculated according to Eq. (1), which was derived in assumption that electron's energy with spin $s = (\uparrow, \downarrow)$ is equal to Fermi energy $E_{F,s}$[10,11]:

$$I_s = \frac{e^2 p_{F,s,\min}^2 a^2 V}{2\pi h^3} \int_0^\infty dk \frac{J_1^2(ka)}{k} F_s(k, D_s, l_s), \quad (1)$$

$$F_s(k, D_s, l_s) = F_s^{bal}(D_s) + F_s^{dif'}(k, D_s, l_s) + F_s^{dif''}(k, D_s, l_s), \quad (2)$$

where $e$ and $l_s$ are electron charge and mean free path, respectively; $a$ is radius of the spot-like contact; $k$ is Fourier image of radial variable $\rho$, which is the coordinate of electron on the contact plane; $J_1(x)$ is Bessel function; $p_{F,s,\min} = \hbar k_{F,s,\min}$ is value of the Fermi momentum, where $k_{F,s,\min}$ is Fermi wavenumber which has to be minimal (min) value between $k_{F,s}^L$ and $k_{F,s}^R$; $F_s(k, D_s, l_s)$ contains integration over $\theta_{c,s}$ which is angle between z-axis and direction of the electron momentum, Fig. 1(a); index $c = L(R)$ determines the contact side. $D_s$ is transmission according standard definition of the quantum mechanics. Transmission usually is a function of the electron wavenumbers, $\theta_{c,s}$, applied voltage $V$ and other possible variables, which defined from specification of the considered problems, where point-contact area can be exchanged by QO, e.g. MTJ with or without NPs, conducting channel with domain wall, etc. In this work, the point-contact area was exchanged by npMTJ and defined as tunneling cell (TC) which was simulated by single and double barrier potential energy profiles [Fig. 1(b) and Fig. 1(c)]. Variable $a$ was defined as a radius of the NP, hence the distance between barriers $d = 2a$.

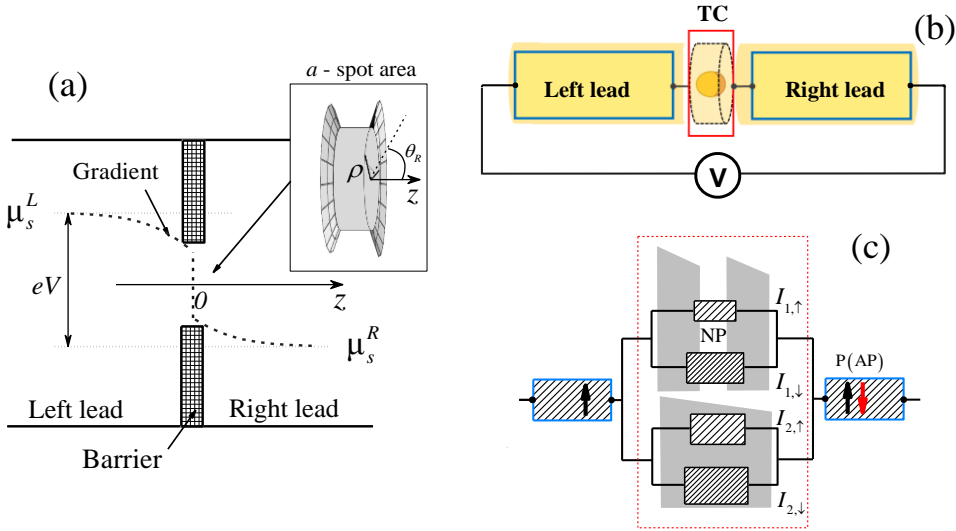

Fig. 1. (a) Schematic plot of the point-like contact, which is considered as an orifice in barrier (*a*-spot model), $\mu_s^{L(R)}$ is chemical potential; (b) TC is a part of the lead-TC-lead system; (c) electric circuit of the lead-TC-lead system.

All integrals inside Eq. (2) are written in relation to variable $\theta_{L,s}$ in order to simplify solution and integral limits. The total current is summation of the both spin components of the charge current $I = I_\uparrow + I_\downarrow$. The complete view of the first and second terms of Eq. (2) was demonstrated by Useinov et al.[10] The first term has a simple form

$$F_s^{bal} = \frac{1}{2\pi} \iint \sin(\theta_{L,s})\cos(\theta_{L,s}) D_s(\theta_{L,s}) d\theta_{L,s} d\varphi \equiv$$
$$\equiv \left\langle \cos(\theta_{L,s}) D_s(\theta_{L,s}) \right\rangle_{\theta_{L,s}},$$

which is responsible for the ballistic and quantum-ballistic transports ($l_s \gg a$). The terms $F_s^{dif\prime}(k, D_s, l_s)$ and $F_s^{dif\prime\prime}(k, D_s, l_s)$ correspond to quasi-ballistic ($l_\uparrow > a$, $l_\downarrow < a$, $l_{\uparrow,\downarrow} \approx a$) and diffusive (Ohmic) regimes of the conduction ($l_s \ll a$). All three terms in Eq. (2) are the solutions of the system for the quasi-classical Green functions with quantum boundary conditions. The last term $F_s^{dif\prime\prime}(k, D_s, l_s)$ contains the gradient terms of the chemical potential nearby contact interface.[11]

In our model, we considered only ballistic case, where $F_s^{bal}$ is dominant single term, while other terms were omitted. In fact, if the barrier thickness of the double barrier subsystem can be reduced to the case where NPs are able to touch both magnetic layers, then ballistic conduction is appeared and most appropriated approach is point-contact model. The consecutive and Kondo-assisted tunneling models and related Coulomb blockade model[3,4] cannot achieve the limit of the ballistic model due to their basic definition.

It is worth to note that $F_s^{bal}$ is not a function of $k$ and $l$. This statement represents an important fundamental quantum property of the electron at nanoscale. In the case when $F_s(k, D_s, l_s) = F_s^{bal}(D_s)$, Eq. (1) can be simplified as follows:

$$\sigma_s = I_s/V = \frac{e^2}{h} \frac{k_{F,s,min}^2 a^2}{2} \left\langle \cos(\theta_{L,s}) D_s \right\rangle_{\theta_{L,s}} \quad (3)$$

Equation 3 is a ballistic part of Eq. (1) and represents simple approach which can be used also for description of the direct electron tunneling in MTJs[5-9]. In case of symmetric nonmagnetic $a$-spot contact $k_F^L = k_F^R$, $\sigma_\uparrow = \sigma_\downarrow$, $D_{\uparrow,\downarrow} \to 1$, $F_s^{bal} = \left\langle \cos(\theta_{L,s}) \right\rangle_{\theta_L} = 1/2$ and $\sigma_\Sigma = \sigma_\uparrow + \sigma_\downarrow = (2e^2/h)(k_F^2 a^2/4)$ which coincides with Sharvin conduction limit[12,13] $\sigma_{Sh}$. Thus, Eq. (3) is some kind of extension of the Sharvin limit, one characterizes the quantum conditions and conduction of the QO which is integrated inside $a$-spot area. The quantum physics of the system can be accounted through analytical or numerical view of the transmission. The conductance [Eq. (3)] is more common case in relation to classical ballistic limit. Hence, its definition in terms of coherent, quantum-ballistic or direct tunneling limits is more appropriated for $D_s \neq 1$. Finally, TMR was defined as following $TMR = (I^P - I^{AP})/I^{AP} \times 100\%$ for TC, and charge currents as $I^{P(AP)} = \sum_s \left( I_{1,s}^{P(AP)} + I_{2,s}^{P(AP)} \right)$ for parallel (P) and anti-parallel (AP) magnetic configurations, respectively. First $I_{1,s}$ and second $I_{2,s}$ current components correspond to the double barrier and single barrier subsystems [Fig. 1(c)].

Furthermore, in order to analyze the solution in range of the complete expression of Eq. (1), all terms in Eq. (2) were simplified in the limit of the non-magnetic symmetric $a$-spot contact:

$$\sigma_\Sigma = 4\sigma_{Sh} \left( \frac{1}{4} - \int_0^\infty \frac{dx}{x} \frac{J_1^2(x)}{1 + (xK)^2 + \sqrt{1+(xK)^2}} \right) \quad (4)$$

where $K = l/a$. In the limit when $K \to \infty$ ($a/l \to 0$), the integral is vanishing in Eq. (4) and $\sigma_\Sigma \to \sigma_{Sh}$. Furthermore, considering that the exact integral's asymptotic for $K \to 0$ ($a/l \to \infty$) is as follows:

$$\lim_{K \to 0} \int_0^\infty \frac{dx}{x} \frac{J_1^2(x)}{1+(xK)^2 + \sqrt{1+(xK)^2}} = \frac{1}{4} - \frac{2}{3\pi} K,$$

it is easy to obtain an exact diffusive solution or Maxwell-Holm limit $\sigma_M$, i.e., $\sigma_\Sigma \to \sigma_M = (8K/3\pi)\sigma_{Sh} = 2a/\rho_V$, where $\rho_V = \hbar k_F/e^2 nl = (e^2 p_F^2 l/3\pi^2 \hbar^3)^{-1}$ is resistivity in volume, and $n = k_F^3/3\pi^2$ is electron density in metals. The dependence of the numerical ratio $\sigma_Z/\sigma_{Sh}$ on $a/l$ is very close to the solution by Wexler with Mikrajuddin's corrections [14,15]

$$\tilde{\sigma}_W/\sigma_{Sh} = \left(\frac{3\pi}{8K}\gamma(K)+1\right)^{-1}, \quad (5)$$

and $\gamma(K) \approx \frac{2}{\pi}\int_0^\infty \exp(-K\cdot x)\mathrm{sinc}(x)dx$, where $\mathrm{sinc}(x) = \sin(x)/x$.

As a result, presented model consists of a quantum-ballistic approach and shows good matching between Sharvin and Maxwell-Holm classical limits.

### 3. Resonant TMR Peaks in npMTJ

Recently, it was theoretically shown that quantized conductance regime is a main reason of the anomalous TMR (suppression or resonant peak-like TMR behaviors) in MTJs with embedded nanoparticles.[1,2] The observed TMR effects are similar to the giant magnetoresistance effects in point-like nano-contacts.[10,16]

The consideration of the system in terms of finite temperature $T$ depends on how corresponding energy $E_F^{NP} = \hbar^2(k^{NP})^2/2m_{eff}$ of NP with $k^{NP}$ is compared to thermal energy $k_BT$, $k_B$ is Boltzmann constant. In range of $k^{NP} < 0.5$ Å$^{-1}$ the thermal energy may have sufficient impact even at low $T$. As a result, when Fermi energy of NP is comparable with $k_BT$, the wave number $k^{NP}$ has the significant margin of the values. The margin $x$ strongly changes the current behavior due to conduction band broadening by following expression:

$$I_s \propto k_{F,s,\min}^2 \int_{X_1}^{X_2} dx \int_0^{\theta_{\min}} \sin(\theta_s)\cos(\theta_s) D_s(\theta_s, k_s^{NP}+x) d\theta_s, \quad (6)$$

where $X_{1(2)} = \mp\sqrt{2m_{eff}k_BT/\hbar^2}$, $m_{eff}$ is effective electron mass. The details of the integration over angle and other parameters can be found in our previous works.[1,2]

In this section, it was assumed that voltage drop is realized only on the TC. As a first-order of the accuracy and simplicity of the npMTJ model, it was proposed that NP has a quasi-continuous energy levels, i.e., the energy level quantization due to possible quantum well states were neglected for the large NPs ($d > 2$ nm). Moreover, it was also considered that the numerical factor which stands out of the integral in Eq. (6) is the same for P and AP current components and it was omitted for TMR simulations by cause of its cancellation. The possible magnetic state and modification of the density of states (DOS) of the NP by reason of exchange interaction between NP and FM leads,[17,18] Kondo physics of NP[19] as well as dispersion of the NPs by size were neglected. Since anisotropies of the DOS dependence of the NP and FM layers due to crystallographic orders were not considered, there is absent of any difference between MTJ with perpendicular and in-plane magnetizations, respectively. According this approach, transmission is independent from the choice of the quantization axis.

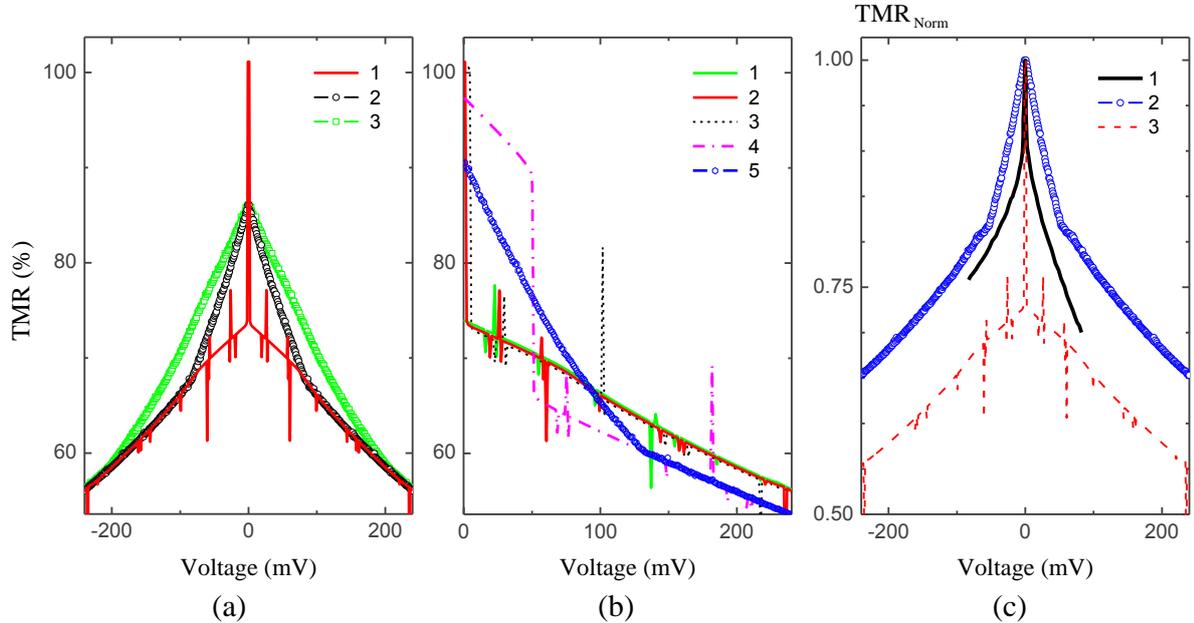

Fig. 2. (a) Resonant TMR peak versus applied voltage at $k^{NP} = 0.4508$ Å$^{-1}$ and its temperature dependence: curves 1-3 correspond to $T = 0.0$ K, 2.5 K and 15 K, respectively; (b) TMR curves 1-5 correspond to $k^{NP} = 0.4512$Å$^{-1}$, $0.4508$Å$^{-1}$, $0.4504$Å$^{-1}$ and $0.445$Å$^{-1}$ at $T = 0.0$ K and $k^{NP} = 0.445$ Å$^{-1}$ at $T = 2.5$ K, respectively. (c) Experimental normalized TMR curve 1 at $T = 2.5$ K [Ref. 4, Fig.2i] and theoretical curves 2 and 3 for $k^{NP} = 0.4508$ Å$^{-1}$ ($T = 1.0$ K, $T = 0$ K). Other parameters are fixed as following: $k_{F,\downarrow}^{L(R)} = 0.421$, $k_{F,\uparrow}^{L(R)} = 1.09$ Å$^{-1}$, $d = 2.6$ nm, $m_{eff}$ in barriers and NP considered as $0.4m_0$ and $0.8m_0$ (where $m_0$ is free electron mass) and for all simulations, the barrier heights and thicknesses are 1.2 eV and 1.0 nm, respectively (color online).

Figure 2 shows resonant peak-like TMR behavior ($d = 2.6$ nm) and its temperature dependence according to Eq. (6). Well-defined resonant TMR peaks were discovered in very narrow range of $k^{NP}$: $0.4504 < k^{NP} < 0.4509$ Å$^{-1}$ for $T \to 0$. In Fig. 2(a), the temperature dependence (curves 1–3) of the peak is shown for $k^{NP} = 0.4508$ Å$^{-1}$ at $T = 0$ K, 2.5 K and 15 K, respectively. Minor resonant peaks are well observed at low $T$ up to 1.5 K, while their complete reduction appeared at $T > 10$ K due to strong thermal averaging.

It is noted that minor resonant peaks are also observed on experiments at low temperatures[3] [Figs. 3(b) and 3(f)]. Moreover, the width of the peak increases, while its amplitude decreases with $k^{NP}$ and $T$ that also is correlated with experimental data[3] [Figs. 2(a) and 2(b)]. The resonant TMR peak becomes to be smoothed under the high temperature which cannot change the resonant conditions. The resonant peak formation at $T \to 0$ is shown in Fig. 2(b), e.g., curve 1 corresponds to $k^{NP} = 0.4512$ Å$^{-1}$ where zero-voltage TMR peak is not yet observed, while curves 2 and 3 with $k^{NP} = 0.4508$ Å$^{-1}$ and $k^{NP} = 0.4504$ Å$^{-1}$ correspond to the well-defined TMR peaks by cause of the beginning of the regime of the quantized conductance. Further decreasing of the $k^{NP}$ value results in step-like TMR behavior (curve 4) which can be smoothed by high temperature (curve 5). It is worth to note that classical (dome-like nonresonant) behavior exists for $k^{NP} > 0.46$ Å$^{-1}$ and the conditions are close to the bulk limit. Figure 2(c) shows experimental and theoretical data comparison of the normalized TMR, $TMR_{Norm} = TMR/TMR_{Max}$. The shape of the TMR-$V$ curve 2 satisfactorily covers experimental data[4] (curve 1).

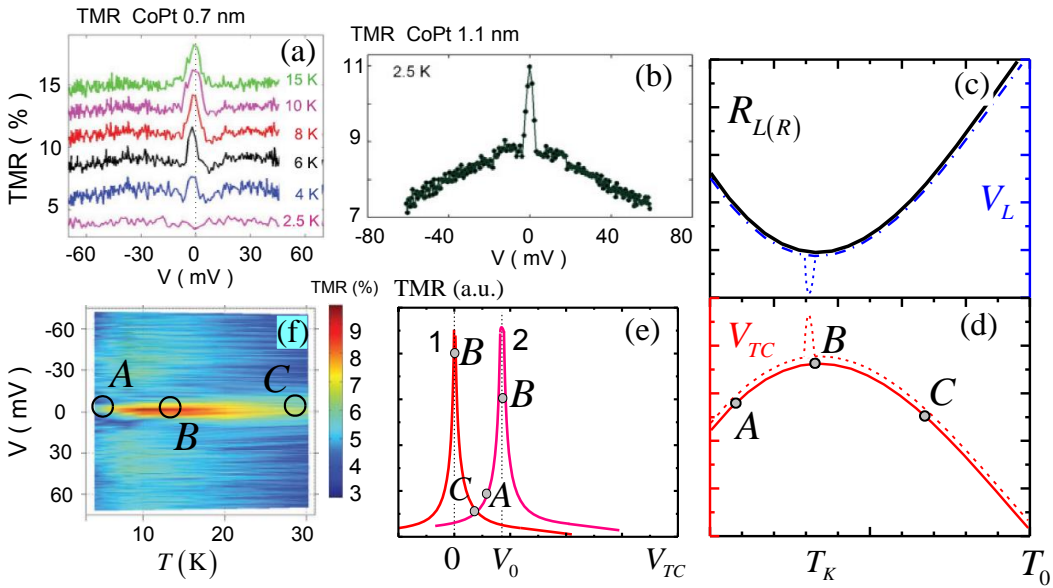

Fig. 3. Panels (a), (b) and (f) show experimental data[3], thickness of the middle layer, responsible for NP formation is 0.7 nm, panels (a) and (f), and 1.1 nm for the panel (b); (c) shows typical resistance behavior (black solid curve) of the leads according Kondo effect; (d) temperature dependence of the voltage drop on the TC from zero to room temperature $T_0$; (e) resonant TMR peak 1 and 2 correspond to the cases without and with $V_0$-shift, respectively; panels (c), (d) and (e) are schematic plots only with arbitrary units (color online).

## 4. *T*-induced TMR Behavior Near Kondo temperature

The approach when applied voltage *V* is fully corresponded to the voltage drop on the TC against conductive electrodes is applicable in the case of the tunnel junction without NP. Its resistance is much larger than resistance of the electrodes (leads) and the voltage drop on the TC $V_{TC}$ is much greater than voltage drop on the leads $V_{L(R)}$. However, considering *T*-induced effect in range of simple qualitative picture with embedded NPs, it was noted that NPs significantly increases the conductance of the TC. Hence, the voltage drop distribution on the whole lead-TC-lead system has to be studied in order to provide more complete model [Fig. 1(b)]. The lead was defined as a part of electrode and magnetic layer, and $V_{L(R)} \leq V_{TC}$.

The temperature dependence of the TMR peak is result of the total resistance change of the lead-TC-lead system, where it is assumed the presence of the Kondo effect for the magnetic leads only in contrary to TC due to the quantum-ballistic regime of the tunneling. It is well known that the ballistic conductance shows lack of dependence on temperature[13]. The thermal heat occurs relatively far away from the *a*-spot area, i.e., on the distance larger than electron mean free path. Experimental setup usually is based on four-probe technique, which allows to compensate the most of lead's resistance impact, however, the result of the measurement is necessary to be estimated more carefully for TCs with nonlinear *I-V*s by the cause of its quantum properties. Figure 3(c) shows schematic Kondo behavior of the lead's resistance $R_{L(R)}$ and related $V_L(T)$. One can see $V_L$ which follows by $R_L$ due to its temperature dependence. It gives voltage drop redistribution in the lead-TC-lead system if the current value is fixed and voltage drop on the whole lead-TC-lead system is $V_{Tot} = V_L + V_{TC} + V_R \approx const$ with the assumption that $V_{Tot} = I(2R_L + R_{TC}) = I \cdot R_{Tot}$ and $V_{L(R)} = I \cdot R_{L(R)}$. Figure 3(d) shows $V_{TC}$ which is inversely proportional to $V_L$ having the maximum around Kondo temperature $T_K$. Thus, *T*-tuning of the $V_{TC}$ can be directly related with the conditions, where the quantized conductance is occurred. Hence, these conditions can be induced in *A*, *B* or *C* points independently [Figs. 3(d) – 3(f)].

Figure 3(e) schematically shows the resonant TMR behavior for two cases where the center of the TMR peak locates directly at zero voltage and the one is shifted by the reason of the intrinsic contact bias $V_0$. For example, the small peak's shift is observed in experiments too,[3] probably by the cause of small $V_0$ [Figs. 3(a) and 3(f)]. One can be also induced by non-equal barrier thicknesses[6]. It is assumed that $V_0$-shift can be compensated according to Kondo effect by varying *T*, $V_L$ and hence $V_{TC}$. The resonant-peak resistance (conductance) of the TC [Fig. 5 in Ref. 1 and Fig. 3(d) in Ref. 2] is much larger (smaller) in comparison to non-resonant one. As a result, $R_{TC} \gg R_L$ and *I-V*s of the lead-TC-lead chain clearly demonstrates quantum behavior of the TC with NP. In other case, when $R_{TC} \geq R_L$ ($V_{TC} \geq V_L$) and $R_{TC} < 2R_L$, the system's *I-V* has the crossover of the classical and quantum regimes of the conductance with reduced TMR value.

It is interesting to note that quantum transport may give dual resonant peak when $V_{Tot}$ is tuned: Fig. 3(f) includes the area showing the yellow-red-yellow-red-yellow zone where voltage is around a few mV and *T* = 20 K. The red color shows highest TMR values. It is important to emphasize that impacts of the $R_{TC}$ and $R_L$ into $R_{Tot}$ are not equivalent, $R_{TC}$ is much more nonlinear than $R_L$. The solid curves in Figs. 3(c) and 3(d) were schematically simplified but since $V_{TC}$ has to have sharp peak (sharp dip for $V_L$), the dashed curves show the realistic case of the resonant behavior. Finally, when the center of the resonant peak directly located at zero voltage, only $B(A) \to C$ occurs, where *A* and *B* are joined and corresponded to the resonant conditions.

It is noted that resonant TMR peak at small $V_{TC}$ arises by the cause of quantized behavior of the spin conduction channels for P and AP magnetic configurations: some conduction channels are still closed for AP configuration and partly opened for P case (i.e., some sections of the permitted $\theta_s$ for AP state are more restricted against P state). At the same time, transmission probability amplitude is coherently suppressed for both P and AP states as a result of the imbalanced matching of the wave functions of the electron on the interfaces. Moreover, the experimentally observed resonant TMR suppression[3,4] means inversed situation, i.e., P state is more restricted against AP but $I^P > I^{AP}$ and $I^P < I^{AP}$ gives positive and negative TMR values, respectively. These effects specify quantum and non-linear properties of the double barrier system.[1,2,5]

## 5. Conclusion

It was demonstrated that direct tunneling plays dominant role and cannot be omitted in the quantum systems with the total tunneling thickness around 5-6 nm. We found that effect of Coulomb blockade can be neglected for the modelling of the anomalous behavior of the TMR in npMTJs. Moreover, non-linear voltage drop redistribution was strongly emphasized and related Kondo effect in the lead-TC-lead system was considered in relation to the temperature dependence of the lead's resistance. Deviation of the voltage drop on the TC provides or suppresses the conditions for the quantized conductance regime (resonant TMR behavior). As a result, temperature dependences of the TMR-*V* curves were simulated at small voltages. The related shapes of the resonant TMR peaks as a function of temperature satisfactorily agreed with reported experimental data.

## Acknowledgments

The work was supported by MOST, Taiwan (grants No.104-2221-E-007-046-MY2 and 103-2112-M-007-011-MY3). A. Useinov acknowledges partial support by the Program of Competitive Growth of Kazan Federal University.